\documentclass[preprint,showpacs,preprintnumbers,amsmath,amssymb]{revtex4}
\usepackage[dvips]{graphicx}
\usepackage{dcolumn}
\usepackage{bm}

\begin{document}


\title{Near-field Optical Spectroscopy and Microscopy of Laterally Coupled Quantum Dots: Bonding and Antibonding States}

\author{Young-Jun Yu}
\author{Haneol Noh}
\author{Gun Sang Jeon}
\affiliation{School of Physics and Astronomy, Seoul National
University, Seoul 151-747, Korea}

\author{Yasuhiko Arakawa}
\affiliation{Research Center for Advanced Science and Technology,
University of Tokyo, Tokyo 153-8505, Japan}

\author{Wonho Jhe}
\email[Corresponding author: ]{whjhe@snu.ac.kr}
\affiliation{School of Physics and Astronomy, Seoul National
University, Seoul 151-747, Korea}


\begin{abstract}

We report on high-resolution photoluminescence (PL) spectroscopic
and microscopic study of {\it laterally coupled} InAs/GaAs
self-assembled quantum dots by using a low-temperature near-field
scanning optical microscope. We have observed slightly split PL
spectra, which are associated with the bonding (symmetric) and
antibonding (antisymmetric) energy states between two coupled
quantum dots, closely located each other as confirmed by spatial
mapping of the PL intensity. The experimental results are in
qualitative agreement with the simple theoretical calculations
based on a two-dimensional potential model. This work may open
the way to a {\it simultaneous} spectroscopy and microscopy study
of laterally coupled quantum dots in a {\it high-density} quantum
dot sample without any articulate sample fabrication.

\end{abstract}

\date{\today}

\maketitle

\newpage


Any coupled single quantum systems, from atoms to molecules to
quantum dots to macro-objects, feature split energy states. For
example, the positively charged hydrogen molecule H$_{2}^{+}$
exhibits energy splitting into the bonding and antibonding states
associated with, respectively, the symmetric and antisymmetric
wave-functions between two neighboring nuclei~\cite{haken}. For
semiconductor quantum dots (QDs), the observation of the split
energies due to coupled QDs or `artificial molecules' has been
also made and the mechanism of an exciton bound to the split
energy states of coupled QDs can be understood by an analogy with
the ionized hydrogen molecules.

Coupled QDs, in particular, have attracted much interest because
the coupling and entanglement of quantized energy states between
neighboring single QDs possess a potential application to quantum
information devices~\cite{lolyd95, bayer01,
gert97,barenco95,borri01}. However, the experimental studies of
coupled QDs have been achieved mostly by using articulately
engineered and prepared samples. For example, `vertically' coupled
QDs were fabricated by vertical alignment of a pair of QD layers
\cite{brian05,goshima05,marek01,gabriel05} and the measured
energy splitting varied from several to several tens of meV
depending on the barrier thickness between the upper and lower
QDs in the bilayer QD
structure~\cite{fujisawa98,bayer01,fonseca98}. The energy
splitting has been also observed in `laterally' coupled QD
molecules, which were specially fabricated by cleaved edge
overgrowth~\cite{gert97} or by combining molecular-beam epitaxy
and atomically precise in-situ
etching~\cite{beirene,krause05,songmuang}.

Self-assembled quantum dots, grown by molecular beam epitaxy, are
usually highly dense with a density larger than
$10^{10}$~cm$^{-2}$. Such a high-density sample of QDs has been
widely studied as a natural candidate for applications to
nano-optical devices. Although the spectroscopic and microscopic
characteristics of self-assembled single QDs have been studied
extensively, coupling between single QDs in such a {\it
high-density} sample has not been investigated, obviously because
it is difficult to discriminate the spectral as well as the
spatial properties of their split energy states. Therefore, it
will be an interesting challenge and significant progress to
study coupling between two adjacent self-assembled QDs because of
its inherent practical potentials for nano and quantum
applications.

In this Letter, we report on the simultaneous spectroscopy and
microscopy study of laterally coupled QDs in a self-assembled
high-density InAs QD sample by using a near-field scanning
optical microscope (NSOM), which has become a powerful tool to
perform laser spectroscopy of single
QDs~\cite{nearapl,saikiprl,yuapl03}. The near-field fiber probe
can measure the high-resolution optical properties of individual
QDs in a high-density ensemble by detecting the spectrally
resolved photoluminescence (PL). Moreover, spatially resolved
mapping of the PL-intensity images can clearly demonstrate that
the bonding (symmetric) and antibonding (antisymmetric) energy
states are associated with the lateral coupling of closely
neighbored single QDs.



The self-assembled InAs QDs were grown on a GaAs substrate,
having an average lateral dimension of 20~nm, a height of 2~nm,
and a density of larger than $10^{10}$~cm$^{-2}$, with deposition
of about two monolayers \cite{todaapl}. The QDs were capped by a
50 nm GaAs layer to reduce the contribution from carrier
diffusion in the barrier layer. For the light source, we used the
Ti:sapphire laser operating near 1.65~eV, which nonresonantly
generates most of charge carriers in the GaAs cap layer. This
excitation laser was coupled to a commercially available
single-mode optical fiber and guided to a chemically etched sharp
fiber tip, on which a 100-nm gold-coated aperture was
fabricated~\cite{jasco}. Such a nanoscale light source generated
by the apertured fiber probe could excite only a few tens of QDs
by the shear-force distance control within several nm from the
capping layer. The resulting PL was collected by the same fiber
so that any loss of spatial resolution due to diffusion could be
minimized \cite{icapl}. The PL signal was then dispersed by a
30-cm long single monochromator with a spectral resolution of
0.3~meV and detected by a liquid-nitrogen-cooled charge coupled
device (CCD) for high signal-to-noise ratio. Both the sample and
the apertured fiber probe were enclosed in a He-flow-type
cryostat and kept at about 10K.



\begin{figure}

\includegraphics[width=0.6\textwidth]{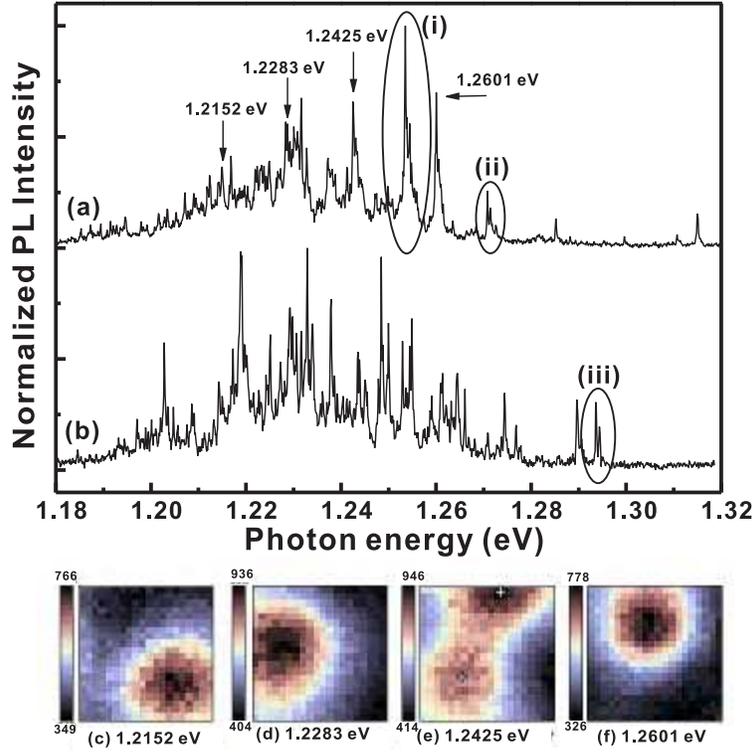}

\caption{Time-integrated PL spectra of high-density InAs single
QDs obtained at two different detection positions of (a) and (b)
on the same sample. PL-intensity NSOM images are shown for the
four selected PL energies in (a), as marked by arrows; (c)
1.2152~eV, (d) 1.2283~eV, (e) 1.2425~eV, and (f) 1.2601~eV. The
scan area is 500~nm~$\times$~500~nm.} \label{fig1}

\end{figure}



Figures~\ref{fig1}(a) and (b) present the  typical
time-integrated, high-resolution PL spectra of InAs QDs collected
during 1~s with an 100-nm apertured probe at two different
positions of the same sample. As can be observed, there are
numerous sharp PL peaks associated with single QDs at a given
position on the high-density sample. The four arrow markers in
Fig.~\ref{fig1}(a) indicate the selected PL energies at which the
PL-intensity NSOM images are also presented; the corresponding
NSOM images are shown in Fig.~\ref{fig1}(c) to (f) for the PL
peaks at (c) 1.2152~eV, (d) 1.2283~eV, (e) 1.2425~eV, and (f)
1.2601~eV.

The NSOM-image data-acquisition process is as follows: (1) The
apertured probe is located on a given sample position and the PL
spectrum is taken in the entire spectral range by the CCD camera
during 1~s. (2) This data taking is repeated sequentially at all
the $20\times20$ pixels of the 500~nm~$\times$~500~nm scanning
area. (3) Each PL-intensity image at a given PL energy is then
processed from the 400 data files, providing the corresponding
energy spectra and spatial images of single
QDs~\cite{yuapl03,yujjap06,eahapl}. We employed the
illumination-collection mode of NSOM to obtain the high-resolution
spatial PL-intensity images, from which one can clearly identify
the isolated single QDs, as shown in Figs.~\ref{fig1}(c), (d) and
(f). On the other hand, as in Fig.~\ref{fig1}(e), one can
sometimes observe the spatial image of PL-intensity exhibiting
two adjacent but independent single QDs having the same energy
within the 0.3 meV resolution limit of the monochromator. The two
cross symbols in Fig.~\ref{fig1}(e) stand for the centers of the
PL images of each QD and the distance between the two crosses is
about 450~nm. This indicates that one can simultaneously observe
the PL-intensity images of two separate single QDs with the same
PL energies in a high-density sample. As is well known, the wave
functions of two individual QDs can be overlapped as the lateral
distance between the QDs is decreased~\cite{gert97, cingolani02}.

\begin{figure}

\includegraphics[width=0.5\textwidth]{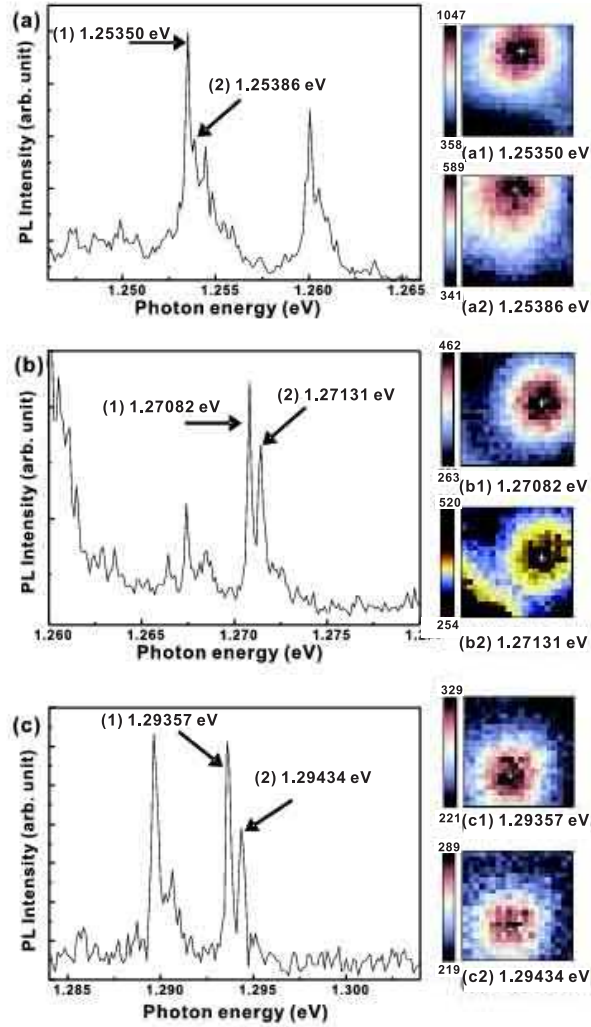}

\caption{Magnified view of the PL spectra shown in (a), (b) and
(c), corresponding to those indicated by the three ellipsoids of
(i), (ii) and (iii) in Figs.~\ref{fig1}(a) and (b), respectively.
The PL-intensity spatial images are also presented at the selected
PL energies. The scan range is 500~nm~$\times$~500~nm.}
\label{fig2}

\end{figure}

In order to study such possible electronic coupling between two
neighboring QDs, we have further investigated the PL spectra which
are slightly split, such as those indicated by the three
ellipsoids in Figs.~\ref{fig1}(a) and (b). The three selected PL
spectra of (i), (ii) and (iii) in Figs.~\ref{fig1}(a) and (b) are
presented in detail in Figs.~\ref{fig2}(a), (b) and (c),
respectively. Each figure shows two slightly separated peaks, as
arrow-marked by (1) and (2), whose corresponding spatial images
are also presented on the right-hand sides as (a1), (a2), (b1),
(b2), (c1) and (c2) in Fig.~\ref{fig2}. In case of
Figs.~\ref{fig2}(a1) and (a2), although the PL energy difference
is only 0.3 meV and thus is almost indistinguishable spectrally,
the spatial PL-intensity images are rather distinguishable; the
separation between the two crosses is about 75~nm (or three
pixels) with each image profile slightly different, depending on
the specific wave-function of each single QDs having different
size, shape, and strain~\cite{saikiprl,yuapl03,yujjap06,eahapl}.
On the other hand, the centers of the PL-intensity images of
Figs.~\ref{fig2}(b1) and (b2) are located at the same position
within one pixel of 25 nm, which limits the spatial resolution of
our measurement. Moreover the two image profiles are very similar
each other with a slightly split energy of 0.49 meV difference.
The same conclusions also hold for the pair of
Figs.~\ref{fig2}(c1) and (c2), with a slightly increased PL-energy
difference of 0.77 meV.

\begin{figure}

\includegraphics[width=0.8\textwidth]{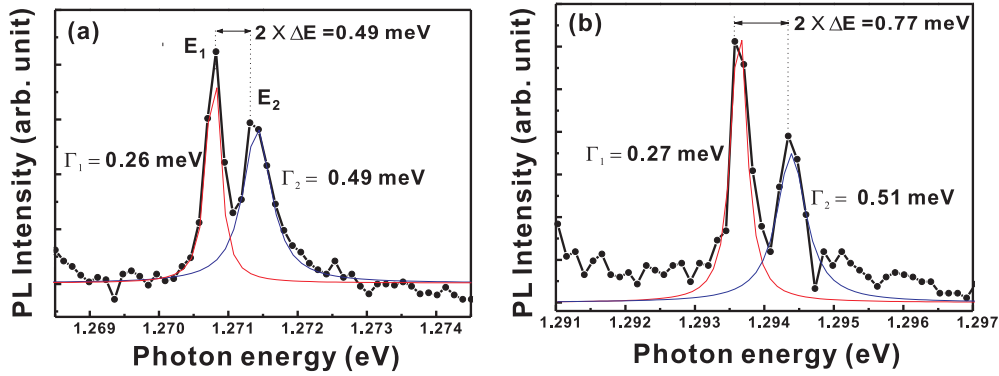}

\caption{Lorentzian fitting results for each PL peaks show the
bonding (left-side curves) and antibonding (right-side curves)
states for the two doublets in Figs.~\ref{fig2} (b) and (c). The
differences between $E_{1}$ and $E_{2}$ are (a) 0.49~meV and (b)
0.77~meV, respectively. The linewidth of the doublet in (a) and
(b) are ($\Gamma$$_{1}$ = 0.26~meV, $\Gamma$$_{2}$ = 0.49~meV)
and ($\Gamma$$_{1}$ = 0.27~meV, $\Gamma$$_{2}$ = 0.51~meV),
respectively.} \label{fig3}

\end{figure}

Figures~\ref{fig3}(a) and (b) present the detailed Lorentzian line
fittings for the doublet PL peaks of Figs.~\ref{fig2}(b) and (c),
respectively. The corresponding PL-energy splittings are 0.49~meV
and 0.77~meV. The associated linewidths are obtained as;
$\Gamma$$_{1}$ = 0.26~meV and $\Gamma$$_{2}$ = 0.49~meV for Fig.
3(a), whereas $\Gamma$$_{1}$ = 0.27~meV and $\Gamma$$_{2}$ =
0.51~meV for Fig. 3(b). From these results, one can observe that
the linewidth of the higher energy PL state ($\Gamma$$_{2}$) is
larger than that of the lower energy state ($\Gamma$$_{1}$),
which is in qualitative agreement with the earlier observation in
Ref.~\cite{gert97} where the broader linewidths were attributed
to acoustic phonon scattering from the upper into the lower state
of a simple two-level system. Note that the values of
$\Gamma$$_{1}$ in Fig.~\ref{fig3} are slightly smaller than the
spectral resolution limit (0.3~meV) of our measurement system,
which indicates that the actual values of $\Gamma$$_{1}$ may be
much smaller than 0.3~meV, as also observed in Ref.~\cite{gert97}.

Johal~\emph{et al.}~\cite{cingolani02} reported that the
edge-to-edge distance between a single dot and its nearest
neighbor in the self-assembled, high-density QD structure is
correlated with the dot size. In other words, the edge-to-edge
distance between two adjacent QDs is decreased with the decrease
of the QD diameter, which indicates that the smaller sized QDs
are more strongly coupled laterally~\cite{cingolani02}. It is
also well known that the higher PL energy is directly associated
with the higher quantization energy of the smaller sized QD.
Consequently, from these two facts, we may deduce that the
coupled QDs are more easily observed at the PL energies higher
than the average PL energy. For our sample, the average lateral
size of single QDs is about 20~nm with the average PL energy of
1.23~eV, as shown in Fig.~\ref{fig1}. Therefore, for the two
coupled QDs having the PL energies of about 1.27~eV and 1.29~eV in
Fig.~\ref{fig2} or Fig.~\ref{fig3}, the size of those coupled QDs
can be assumed to be smaller than
20~nm~\cite{yuapl03,japQDsize,aplQDsizebr,prbQDsize}. Moreover,
it can be justified that the doublet PL peaks in Figs.~\ref{fig2}
(b) and (c) are not due to the internal energy states of a single
QD, but result from the coupled states between a pair of
interacting adjacent small-sized QDs, because the typical energy
difference between the quantized energy states of a single InAs
QD is several tens of meV~\cite{japQDsize,aplQDsizebr,prbQDsize},
much larger than the energy splitting of about 1~meV in
Fig.~\ref{fig2}. It should be emphasized again that the results
in Fig.~\ref{fig2}(a) look like coupled QDs, but the PL-intensity
images clarify that they are due to two uncoupled QDs having the
similar energies, not from two coupled QDs located closely nearby.

Let us now make a qualitative description of the energy splitting
in the laterally coupled QDs by a simple numerical analysis. In
previous works, the doublet PL features such as those in
Figs.~\ref{fig2} and \ref{fig3} were explained as due to the
excitonic emission of the bonding (symmetric) and antibonding
(antisymmetric) states that result from the coupled ground states
of independent single QDs~\cite{gert97,cingolani02}; the energies
of the bonding~($E_{1}$) and antibonding~($E_{2}$) states were
expressed as $E_{1,2} = E_{0}~\mp~\Delta E$ by the first
perturbation theory. Here E$_{0}$ is the unperturbed energy of
the individual QD and $\Delta E$ is the matrix element such that
the coupling-induced energy splitting is $2\times \Delta
E$~\cite{gert97}. Since the energy splitting ($2\times \Delta E$)
of the doublet states depends on the separation between two QDs
as well as their size \cite{bayer01,gert97}, one may consider
that the larger QDs with the lower PL energies are coupled in
Fig.~\ref{fig3}(a), whereas the smaller QDs with the higher PL
energies are coupled in Fig.~\ref{fig3}(b), as discussed before.
Moreover, the exact distance between two coupled QDs cannot be
directly measured because their separation is beyond the available
spatial resolution limit of $25$ nm. Nonetheless, for qualitative
justification, we have performed a simple model calculation to
estimate the distances between two coupled QDs.

In our analysis, each QD is modeled as a two-dimensional circular
potential-well of diameter $D$,
\begin{equation}
V(r) = \left\{
\begin{array}{cl}
- V_0 & \hbox{ for } r < D/2,\\
0 & \hbox{ for } r > D/2.
\end{array}
\right.
\end{equation}
For a single QD located at the origin, the normalized wave
function of its electronic ground state is given by,
\begin{equation}
\psi_0(r) = \left\{
\begin{array}{cl}
\mathcal{N} K_0 (\kappa D/2) J_0(k r) & \hbox{ for } r < D/2,\\
\mathcal{N} J_0 (k D/2) K_0(\kappa r) & \hbox{ for } r > D/2,
\end{array}
\right.
\end{equation}
where $k^2 \equiv 2 m_{\rm eff} (E_0 + V_0)/\hbar^2 $, $\kappa^2
\equiv 2 m_{\rm eff} |E_0|/\hbar^2$ ($m_{\rm eff}$ is the
effective mass of electron), $J_0$ $ (K_0)$ is the zeroth-order
(modified) Bessel function of the first (second) kind, and
$\mathcal{N}$ is a normalization constant. Here the ground-state
energy $E_0$ is determined by the lowest energy such that $k
J_1(kD/2) / J_0(kD/2) = \kappa K_1(\kappa D/2) / K_0(\kappa
D/2)$. The bonding and antibonding states are then obtained by
direct diagonalization of the Hamiltonian which describes two
single QDs separated by the edge-to-edge distance of $d$. The
diagonalization is performed in the Hilbert space spanned by the
ground states of each QD. For our simulation, we used $V_0=0.6$
eV and $m_{\rm eff}/m_0 = 0.023$ ($m_0$ is the bare electron
mass)~\cite{bayer01,marek01}. The diameter $D$ of similar single
QDs associated with each PL peaks in Fig.~\ref{fig3} can be also
estimated from their PL energies; the PL energies of 1.271 eV and
1.294 eV correspond to the diameters $D$ of $15.1$ and $13.4$ nm,
respectively, because the average PL energy of 1.23 eV is
attributed to the average diameter of $D = 20$ nm.

\begin{figure}

\includegraphics[width=0.6\textwidth]{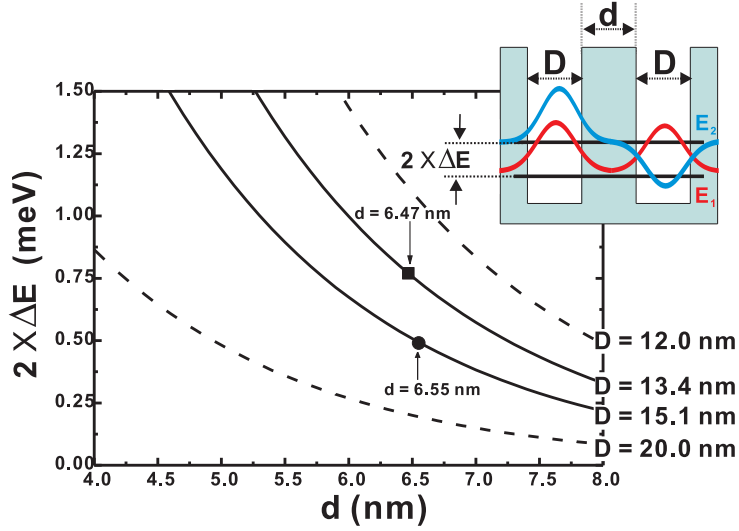}

\caption{Calculated results of correlation between the energy
splitting (2$\times$$\Delta$E) and the edge-to-edge distance ($d$)
of coupled QDs for various QD diameters ($D$ $=$ 12.0~nm,
13.4~nm, 15.1~nm and 20~nm). The inset is the schematic diagram
of the split bonding (symmetric) and antibonding (antisymmetric)
ground-states of the coupled QDs.} \label{fig4}

\end{figure}

Figure~\ref{fig4} shows the computed energy splittings $2\times
\Delta E$ as a function of the edge-to-edge distance $d$ for
single QDs of various diameters $D$. The numerical curves then
enable one to estimate the effective barrier width $d$ (as shown
in the inset of Fig.~\ref{fig4}) from the experimental values of
the energy splitting. We find that the coupled QDs in
Figs.~\ref{fig3}(a) and (b) correspond, respectively, to $d=6.55$
and $6.47$ nm, both smaller than the QD sizes, as expected.
Therefore, by combining Figs.~\ref{fig2}, \ref{fig3} and
\ref{fig4}, one may be able to investigate the doublet PL spectra
of the bonding and antibonding states, as well as to estimate the
separation of two coupled InAs QDs embedded in a 50~nm GaAs
capping-layer.


In conclusion, we have employed the PL NSOM method to obtain
spatially and spectrally resolved PL-intensity images due to {\it
laterally coupled} single QDs in a self-assembled, high-density
QD sample. In particular, the high-resolution PL spectra and the
PL-intensity images exhibit the bonding and antibonding states
associated with two coupled, closely adjacent QDs. By a simple
model calculation, we have also estimated the separations between
coupled QDs. This work may open the possibility to investigate
the coupling interactions between closely neighboring QDs in a
high-density QD sample for quantum optics or quantum information
applications with the simultaneous high spectroscopic as well as
microscopic resolution.

\begin{acknowledgements}
This work was supported by the Korean Ministry of Science and
Technology.
\end{acknowledgements}









\end{document}